\documentclass[doublecol]{epl2} 
\newcommand{\bea}{\begin{eqnarray}}    
\newcommand{\eea}{\end{eqnarray}}      
\newcommand{\be}{\begin{equation}}
\newcommand{\ee}{\end{equation}}
\newcommand{\bef}{\begin{figure}}
\newcommand{\eef}{\end{figure}}

\def\spose#1{\hbox to 0pt{#1\hss}}
\def\ltapprox{\mathrel{\spose{\lower 3pt\hbox{$\mathchar"218$}}
\raise 2.0pt\hbox{$\mathchar"13C$}}}
\def\gtapprox{\mathrel{\spose{\lower 3pt\hbox{$\mathchar"218$}}
\raise 2.0pt\hbox{$\mathchar"13E$}}}
\def\inapprox{\mathrel{\spose{\lower 3pt\hbox{$\mathchar"218$}}
\raise 2.0pt\hbox{$\mathchar"232$}}}

\title{Persistent fluctuations  in the distribution of galaxies
from the Two degree Field Galaxy Redshift Survey}
\shorttitle{Persistent fluctuations  in the distribution of galaxies}

\author{Francesco Sylos Labini\inst{1,2} 
\and Nikolay L. Vasilyev \inst{3} 
\and Yurij V. Baryshev \inst{3}
}
\shortauthor{F. Sylos Labini \etal}

\institute{                    
  \inst{1} Museo Storico della Fisica e Centro Studi e
  Ricerche Enrico Fermi, - Piazzale del Viminale 1,
  00184 Rome, Italy \\
  \inst{2}  Istituto dei Sistemi 
  Complessi CNR, - Via dei Taurini 19, 00185 Rome, Italy \\
 \inst{3} Institute of Astronomy, St.Petersburg State
  University - Staryj Peterhoff, 198504, St.Petersburg, Russia
}
\pacs{98.80.-k}{Cosmology}
\pacs{05.40.-a}{Fluctuations phenomena in random processes}
\pacs{02.50.-r}{Probability theory, stochastic processes, 
and statistics}

\abstract{ We apply the scale-length method to several three
  dimensional samples of the Two degree Field Galaxy Redshift Survey.
  This method allows us to map in a quantitative and powerful way
  large scale structures in the distribution of galaxies controlling
  systematic effects.  By determining the probability density function
  of conditional fluctuations we show that large scale structures are
  quite typical and correspond to large fluctuations in the galaxy
  density field. We do not find a convergence to homogeneity up to the
  samples sizes, i.e. $\approx 75$ Mpc/h.  We then measure, at scales
  $r\ltapprox 40$ Mpc/h, a well defined and statistically stable
  power-law behavior of the average number of galaxies in spheres,
  with fractal dimension $D=2.2 \pm 0.2$.  We point out that standard
  models of structure formation are unable to explain the existence of
  the large fluctuations in the galaxy density field detected in these
  samples. This conclusion is reached in two ways: by considering the
  scale, determined by the linear perturbation analysis of a
  self-gravitating fluid, below which large fluctuations are expected
  in standard models and through the determination of statistical
  properties of mock galaxy catalogs generated from cosmological
  N-body simulations of the Millenium consortitum}
\begin{document}

\maketitle

\section{Introduction}

 In the past twenty years observations have provided growing evidences
 that galaxy distribution is organized in a complex network of
 structures and voids \cite{devac1970,cfa2,colless01,york}.  Despite
 the fact that large scale galaxy structures, of size of the order of
 several hundreds of Mpc/h\footnote{We use $H_0=100 h$ km/sec/Mpc for
   the value of the Hubble constant.}, have been observed to be the
 typical feature of the distribution of visible matter in the local
 universe, the statistical analysis measuring their properties has
 identified a characteristic scale which has only slightly changed
 since its discovery fourthy years ago in angular catalogs. This
 scale, $r_0$, was measured to be the one at which fluctuations in the
 galaxy density field are about twice the value of the sample density
 and it was indeed determined to be $r_0 \approx$ 5 Mpc/h in the Shane
 and Wirtanen angular catalog \cite{tk69}. Subsequent measurements of
 this scale --- see e.g. \cite{dp83,davis88,park,benoist,norbergxi01,
   norbergxi02,zehavietal02,zehavietal04} --- found a similar value,
 although in several samples larger values of $r_0$ have been found
 (i.e. $r_0 \approx 6-12$ Mpc/h).  This variation was then ascribed to
 a luminosity dependent effect --- see e.g.
 \cite{davis88,park,benoist,zehavietal02}.

However, recently in a CCD survey of bright galaxies within the
Northern and Southern strips of the 2dF Galaxy Redshift Survey
(2dFGRS) \cite{colless01} conclusive evidences where found that there
are fluctuations of the order $ \sim 30\%$ in galaxy counts as a
function of apparent magnitude \cite{busswell03} (see also
\cite{frith03,frith06} for similar observations in other galaxy
samples).  Further since in the angular region toward the Southern
galactic cap (SGC) a deficiency, with respect to the Northern galactic
cap (NGC), in the counts below magnitude $\sim 17$ (in the $B$ filter)
was found, persisting over the full area of the APM and APMBGC
catalogs, this would be an evidence that there is a large void of
radius of about $150$ Mpc/h implying that there are spatial
correlations extending to scales larger than the scale detected by the
2dFGRS correlation function \cite{norbergxi01,norbergxi02}. Indeed, by
considering the two-point correlation function, and thus by
normalizing the amplitude of fluctuations to the estimation of the
sample density, the length-scale $r_0 \approx 6-8$ Mpc/h was derived
\cite{norbergxi01,norbergxi02}.

Structures and fluctuations at scales of the order of 100 Mpc/h or
more are at odds with the prediction of the concordance model of
galaxy formation \cite{busswell03,frith03,frith06}, while the small
value of the correlation length is indeed compatible.
In what follows we try to clarify this puzzling situation, i.e. the
coexistence of the small typical length scales measured by the
two-point correlation function analysis with the large fluctuations in the
galaxy density field on large scales as measured by the simple galaxy
counts. Because of the difference in the counts amplitude, and thus in
the sample density between the NGC and the SGC samples, the estimation
of the sample density is not stable and thus one must critically
consider the significance of the normalization of fluctuations
amplitude to the estimation of the sample density as used in the
correlation analysis employed to measure the length scale $r_0$.

More generally the problem of the statistical characterization these
structures in a finite sample, of volume $V$ containing $M$ galaxies,
can be rephrased as the problem of measuring volume averaged
statistical quantities. The basic issue concerns whether these are
meaningful descriptors, i.e. whether they give or not stable
statistical estimations of ensemble averaged quantities \cite{book}.
In general it is assumed that galaxy distribution is an ergodic
stationary stochastic process \cite{book}, which means that it is
statistically translationally and rotationally invariant, thus
avoiding special points or directions.  Stationary stochastic
distributions satisfy these conditions also when they have zero
average density in the infinite volume limit \cite{book}.  The
assumption of ergodicity implies that in a single realization of the
microscopic number density field $n(\vec{r})$ the average density
$n_0$ in the infinite volume is well defined and equal to the ensemble
average density \cite{book}.  The constant $n_0$ is strictly positive
for homogeneous distributions and it is zero for infinite
inhomogeneous ones \cite{book}. The infinite volume limit must be
considered in the definition of probabilistic properties, but in
physical systems one is concerned only with finite volumes and
statistical determinations.  For inhomogeneous distributions, in a
finite sample, the estimation of the average mass density gives a
large relative error with respect to the ensemble value and it is thus
systematically biased \cite{book}.  This situation occurs as long as
the sample size is smaller than the scale $\lambda_0$ at which the
distribution turns to homogeneity, i.e.  beyond which density
fluctuations are small \cite{book}.  In the finite sample analysis it
is then necessary to study the conditional scaling properties of
statistical quantities, by an analysis of fluctuations and
correlations which explicitly considers whether a distribution can be
or not homogeneous. Before turning to the description of the methods
employed to study galaxy distributions and mock galaxy samples we
discuss the properties of the samples considered.

\section{The Data}
\label{samples_sdss}

The Two degree Field Galaxy Redshift Survey (2dFGRS)\footnote{\tt
  http://www.mso.anu.edu.au/2dFGRS/} \cite{colless01} measured
redshifts for more than $220,000$ galaxies in two strips in the
southern galactic cap (SGC) and in northern galactic cap (NGC).  The
median redshift is $z\simeq0.1$ and the apparent magnitude corrected
for galactic extinction in the $b_J$ filter is limited to
$14.0<b_J<19.45$.  The selection of the samples used in the analysis
discussed below is described in \cite{2df_paper}.  To avoid the effect
of the irregular edges of the survey we selected two rectangular
regions whose limits are: for SGC $84^\circ\times 9^\circ$
($-33^\circ<\delta<-24^\circ$, $-32^\circ<\alpha<52^\circ$), and for
NGC: $60^\circ\times 6^\circ$ ($-4^\circ<\delta<2^\circ$,
$150^\circ<\alpha<210^\circ$).  We construct Volume Limited (VL)
samples, which are unbiased for the observational selection effects
due to the limit in apparent magnitude \cite{zehavietal02}. To this
aim we computed the metric distance $R(z)$ with parameters
$\Omega_M=0.3$ and $\Omega_\Lambda=0.7$ (i.e. the concordance model)
and we determined absolute magnitudes $M$ using K-corrections from
\cite{madgwick02}. Two couples of VL samples, in each galactic cap,
are identified by (i) 100 Mpc/h $< R <$ 400 Mpc/h and -19.0 $< M<$
-20.8 (SGC400 and NGC400) and (ii) 150 Mpc/h $< R <$ 550 Mpc/h and
-19.8 $< M<$ -21.2 (SGC550 and NGC550). Each sample contains about
$N\approx 2\div 3 \cdot 10^4$ galaxies \cite{2df_paper}.

\section{Statistical methods}

\begin{figure*}
\begin{center}
%\onefigure[scale=0.3]
\onefigure[scale=1]{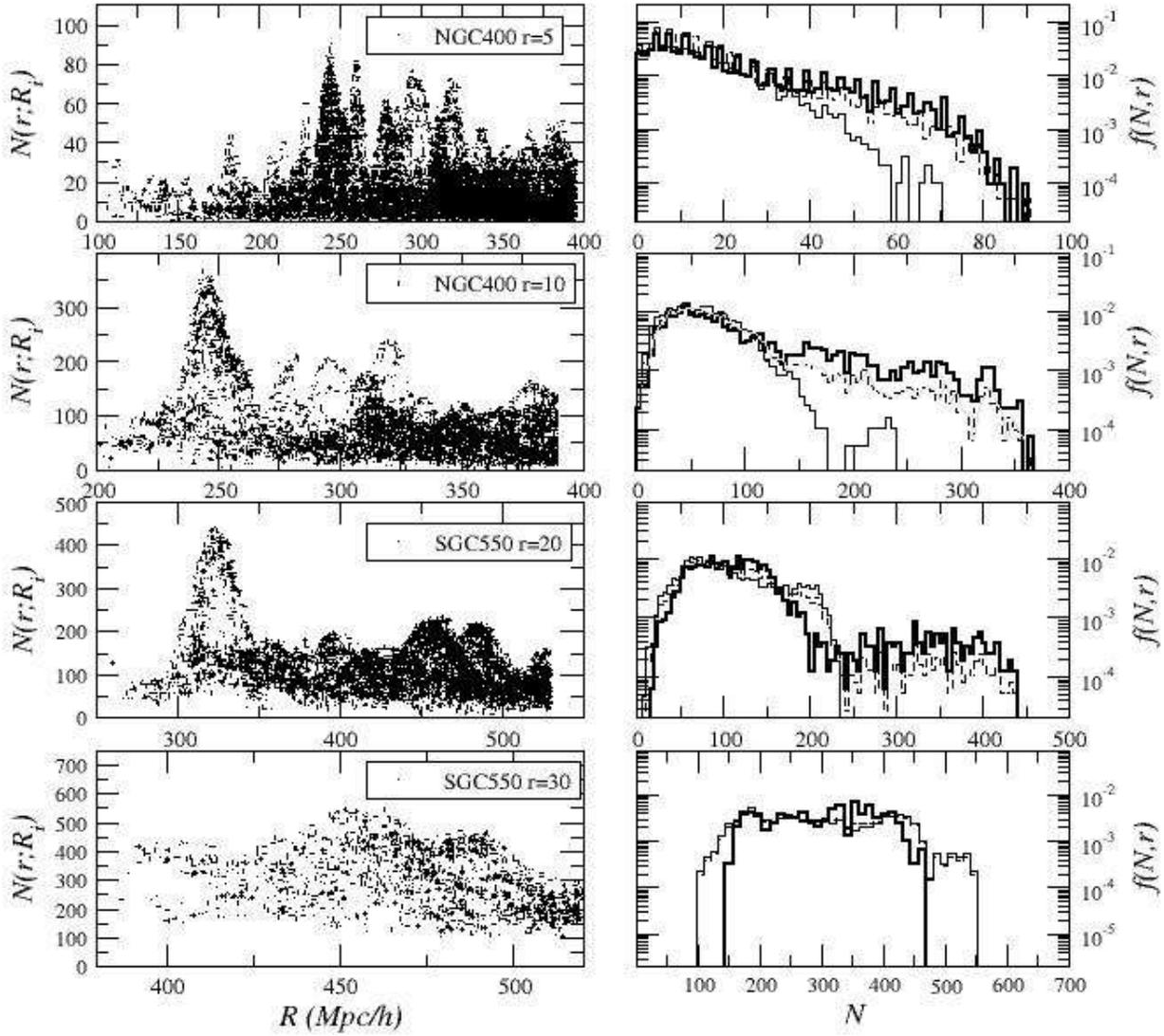}
\caption{
{\it Left panels}: From top to bottom the SL analysis for
  the different 2dFGRS samples and $r=5,10$ Mpc/h (NGC400) and
  $r=20,30$ Mpc/h (SGC550).    {\it Right panels}:
  Probability density function $f(N,r)$ of $N(r;R_i)$ in the whole
  sample (thick solid line) and in two non-overlapping sub-samples with
  equal volume (each half of the sample volume) at small (think solid line)
  and large (dashed line) $R$. }
\label{fig_sl}
\end{center}
\end{figure*}

The scale-length (SL) analysis \cite{paper1} consists in the
determination of the number $N(r;R_i)$ of galaxies in spheres of
radius $r$, centered on the $i^{th}$ galaxy at the radial
distance $R_i$ from the observer. When this is averaged over the whole
sample it gives an estimate of the average conditional number of
galaxies in spheres of radius $r$ \cite{paper1,book}
\be
\label{eq1} 
\overline{N(r)} = \frac{1}{M(r)} \sum_{i=1}^{M(r)} N(r;R_i) \;, 
\ee
where the sum is extended to the $M(r)$ galaxies whose distance from
the boundaries of the sample is smaller or equal to $r$.  In this way
when $r$ growths $M(r)$ decreases with $r$ because only those galaxies
for which the sphere is fully included in the sample volume are
considered as centers \cite{book}.  In addition when $r$ is large
enough only a part of the sample is explored by the volume average
\cite{book,paper1}.  Thus for large sphere radii $M(r)$ decreases and
the location of the galaxies contributing to the average in
Eq.\ref{eq1} is mostly at radial distance $\sim [R_{min}+r$,
  $R_{max}-r]$ from the radial boundaries of the sample at $[R_{min}$,
  $R_{max}]$.  By using these boundary conditions Eq.\ref{eq1} gives
the so-called full-shell estimator \cite{book,kerscher}. This has the
advantage to make the weakest a-priori-assumptions about the
properties of the distribution outside the sample volume.
Indeed one may use incomplete spheres, by counting the galaxies inside
a portion of a sphere and by weighting this for the corresponding volume
\cite{kerscher}. However this method implicitly assumes
 that what is inside the incomplete sphere is  a
statistically meaningful estimate of the whole spherical volume. This
is incorrect when a distribution presents large fluctuations. For
example in the part of a spherical volume which lies outside the
sample boundaries there can be an empty region or a large
scale structure: in this situation the weighted estimate is biased
\cite{book}.
In a finite sample, together with the average given by
  Eq.\ref{eq1}, one may determine amplitudes of density fluctuations
by measuring their variance $\overline{\delta(r)^2} \equiv
[\overline{N(r)^2} - \overline{N(r)}^2]/ \overline{N(r)}^2 \sim 1,$
where the last equality holds for inhomogeneous distributions
(e.g. for a fractal $\overline{N(r)} \sim r^D$ and $D<3$) and it means
that fluctuations are persistent \cite{book}.  In such a situation,
because, of the strong correlations, the Central Limit Theorem does
not hold and the probability density function (PDF) of fluctuations
does not generally converge to a Gaussian function as for homogeneous
ones \cite{book}, where $\overline{\delta(r)^2} \ll 1$ \cite{book}.

The SL analysis of two 2dFGRS samples (Fig.\ref{fig_sl}) shows large
density fluctuations in the locations corresponding to large scale
structures. Large scale structures transversely cross the NGC400
volume at about 250, 260, 290 and 320 Mpc/h of thickness of about 30
Mpc/h.  When the sphere radius $r$ is increased from 5 to 10 Mpc the
most prominent structure is the one at about $250$ Mpc/h.  This is due
to the geometrical selection effect previously discussed.  In the
SGC550 sample the situation is similar to the NGC400 case, except for
the fact that the radial distances corresponding to the large
variations (i.e. structures) in $N(r;R_i)$ are different.  We note
that the same structures we observe in Fig.\ref{fig_sl} have been also
identified with different methods \cite{eke04,einasto}.

Thus, galaxy distribution in these samples are dominated by several
large scale structures which cross their volumes.  These structures
are typical, i.e. they are detected at different radial distances and
in two different sky areas of the 2dFGRS.  They correspond to large
density fluctuations, i.e.  large variations of $N(r;R_i)$.  For
largest sphere radius we considered is $r\approx 40$ Mpc/h we find
fluctuations of order four in $N(r;R_i)$. This implies that
$\lambda_0>40$ Mpc/h.

\section{Convergence to homogeneity ?}

\begin{figure}
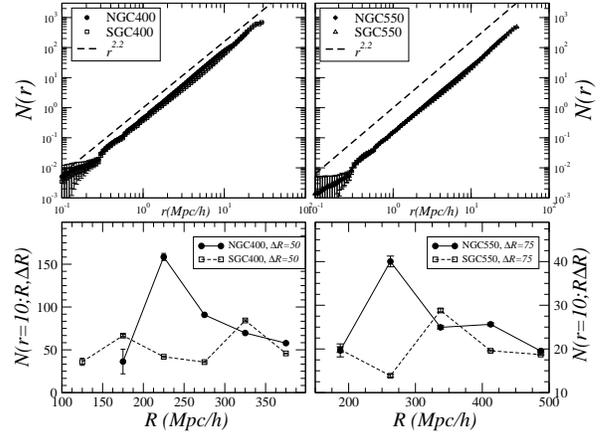

\onefigure[scale=0.3]{Fig2.eps}
\caption{ {\it Upper panels}: In the left panel it is shown the
    average number of points in spheres of radius $r$ around a galaxy
    for the samples SGC400 and NGC400, while in the right panel for
    SGC550 and NGC550. A reference line with a slope 2.2 is reported
    with the sample amplitude in both panels. The different amplitude
  in the two pairs of samples is ascribed to the different limits in
  absolute magnitude \cite{book}.  {\it Bottom panels}: In the left
  panel we show average value of $N(r;R_i)$ with $r=10$ Mpc/h in bins
  of $\Delta R=50$ Mpc/h while in the right panel the same for $r=10$
  Mpc/h and $\Delta R=75$ Mpc/h.}
\label{fig_NR}
\end{figure}

We can now use the data obtained by the SL analysis to investigate
whether there is a convergence to homogeneity at some large scales
$r >  40$ Mpc/h.  This is achieved by dividing the whole range of
radial distances in bins of thickness $\Delta R$ centered at $R$,  
and computing in each bin the
average $\overline{N(r;R,\Delta R)}$ of $N(r;R_i)$ at fixed $r$ 
(Fig.\ref{fig_NR}).  To this aim we used
 small radii $r=5,10$ Mpc/h in order to avoid substantial overlap in
space between neighboring bins in radial distance.  We expect that, if
the distribution converges to homogeneity at a scale $\lambda_0$,
correspondingly $\overline{N(r;R,\Delta R>\lambda_0)}$ does not show
large fluctuations as a function of $R$.
On the other hand for the largest radial bin chosen $\Delta R=75$
Mpc/h the measurements in bins centered at different $R$ wildly
scatters in the SGC and NGC samples, i.e. their values are outside the
statistical error bars.  These results show that structures, leading
to persistent fluctuations up the largest scales sampled by this
catalog, have an amplitude which is incompatible with homogeneity at
scales $\lambda_0 \le \Delta R = 75$ Mpc/h.  A smooth
redshift-dependent correction (i.e. galaxy evolution) would yield, at
fixed $R$, to the same corrections in the SGC and NGC: thus the
fluctuations detected cannot be an artifact of such an effect.

The frequency distribution of conditional fluctuations gives 
an estimation of the PDF  $f(N,r)$: this  is computed at fixed
sphere radius $r$  (Fig.\ref{fig_sl}). In all cases $f(N,r)$ is
non Gaussian and statistically stable, i.e.  it does not change when
it is computed in the whole sample or in two non-overlapping
sub-samples with equal volume at small and large radial distance. The
tail of $f(N,r)$ for large values of $N$ is instead affected by the
different fluctuations in the different sub-volumes.  The trend is
obvious: larger the fluctuations of $N(r;R_i)$ more extended toward
large $N$ values is the tail of $f(N,r)$. In each sub-sample, for the
largest sphere radius $r$ we find that $f(N,r)$ is systematically
distorted with respect to smaller sphere radii. This is due to fact
that, for large sphere radii, the volume average cannot explore
properly the full sample because of the geometrical selection effect
present in the determination of $N(r;R_i)$ and it is dominated by only
few structures.  The determination of the whole-sample average
statistics, i.e. Eq.\ref{eq1}, provides a meaningful statistical
quantity as the PDF in all samples is reasonably statistically stable.
We find $\overline{N(r)} \propto r^D$ with $D=2.2\pm0.2$ up to $r \sim
40$ Mpc/h (Fig.\ref{fig_NR}) in agreement with previous determinations
by \cite{2df_paper}.

\section{Estimation of the standard two-point correlation function}

The estimator of the two-point correlation function can be written as
\cite{book} 
\be
\label{xiestim} 
\overline{\xi(r)} +1 = \frac{\overline{dN(r)}} {4\pi r^2 dr} \cdot
\frac{1} {n_S} \;, \ee
where $n_S$ is the sample density.  If $\overline{N(r)} \propto r^D$
then $\overline{\xi(r)}$ has the following features: (i) its amplitude
is proportional to the sample size and (ii) it shows a break from a
power-law at a scale of the order of the sample size. The amplitude of
$\overline{\xi(r)}$ is then a ratio between a local and a global
quantity ($n_S$). The former one can be estimated for instance as
$n_S=N/V$. When $V$ is spherical of radius $R_s$ we get that
$\xi(r_0)=1$ for $r_0 = (D/6)^{1/(3-D)} R_s$. The geometry of the
2dFGRS samples is a spherical portion for which the radius of the
maximum sphere fully enclosed is about $R_s\approx 40$. Given that $D
\approx 2$, we get $r_0\approx 10$ Mpc/h, which is approximately the
value obtained by \cite{norbergxi01,norbergxi02}. The normalized mass
variance is equal to unity at approximately the same scale
\cite{book}. A more detailed discussion of the determination of the
two-point correlation function can be found in
\cite{2df_paper,nslb08}.

\section{Comparisons with standard models of galaxy formation}

In cosmological structure formation cold dark matter models
\cite{peacock} gravitational collapse firstly forms non-linear
structures (i.e. large fluctuations) at small scales and then larger
and larger scales become non-linear.  The theoretical homogeneity
scale $\lambda^m_0$ identifying the range of distances where large
inhomogeneities are formed, can be defined to such that the
unconditional relative mass variance in spheres is
$\sigma^2(\lambda^m_0) =1$ \cite{pee80}. Thus from the time dependence
of the power spectrum in the linear perturbation analysis of the
self-gravitating fluid equations it is possible to derive the
time-dependence of $\lambda^m_0(t)$ \cite{pee80}.  By normalizing the
initial amplitude of density fluctuations to the Cosmic Microwave
Background Radiation (CMBR) anisotropies it is found that at the
present time, in the concordance model, $\lambda_0^m \approx 10$ Mpc/h
\cite{springel05,spergel,cdm_theo}.

\begin{figure}
\onefigure[width=80 mm]{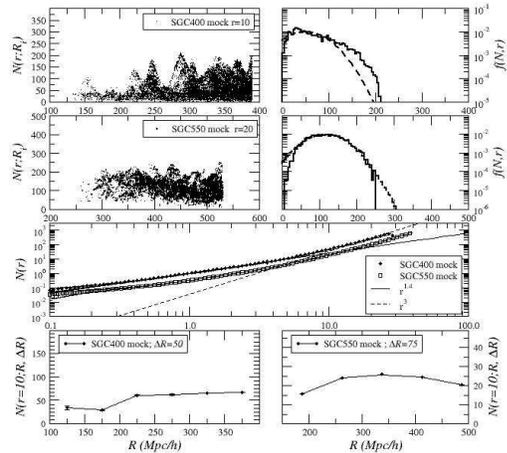}
\caption{ From top to bottom: The SL analysis for $r=10,20$ Mpc/h
  (left) for the mock samples and their PDF (right). The dashed line
  is a the best fit with a Gaussian function.  {\it Fifth panel}:
  Average number of points in spheres of radius $r$ around a
  galaxy. The reference lines have slopes $D=1.2$ (in real space) for
  $r<10$ Mpc/h and $D=3$ for $r>10$ Mpc/h.  {\it Bottom panels}:
  (left) average value of $N(r;R_i)$ with $r=10$ Mpc/h in bins of
  $\Delta R=50$ Mpc/h; (right) the same for $r=10$ Mpc/h and $\Delta
  R=75$ Mpc/h.}
\label{fig_mock_sl}
\end{figure}

This estimation is in agreement with results of cosmological N-body
simulations which are used to study the non-linear regime for
$r<\lambda_0^m$.  Here we considered the cosmological simulations
performed by the Millennium project which are the largest ones
performed until now \cite{springel05}.  Amount of dark matter and
cosmological parameters are given in agreement with the standard
concordance models.  Dark matter simulations have about $10^{10}$
particles and galaxies are identified according to semi-analytics
models of galaxy formation \cite{cronton06}.  We used a mock galaxy
catalog with about 9 millions objects, where absolute magnitudes of
mock galaxies can be transformed in the same filter $b_J$ of the
2dFGRS \cite{cronton06}.  We have cut samples with almost same
geometry, number of objects and limits in magnitude and distance as
the 2dFGRS samples.  Note that we computed $N(r;R_i)$ where $R$ is the
real-space position of each object with respect to the observer: in
redshift space the dimension for $r<10$ Mpc/h is slightly different
\cite{nslb08}.  We find (Fig.\ref{fig_NR}) that $\overline{N(r)} \sim
r^{1.2}$ for $r<10$ Mpc/h and $\overline{N(r)} \sim r^3$ for $r>10$
Mpc/h. The PDF of conditional fluctuations rapidly converges to a
Gaussian for $r>10$ Mpc/h and it is statistically stable
(Fig.\ref{fig_mock_sl}). Correspondingly $N(r;R_i)$ shows different
and more quiet fluctuations than the real data.  
Note that at scales $r>10$ Mpc/h density fluctuations in the dark
  matter field are in the linear regime and thus the understanding of
  biasing in that case is simple.  Indeed, according to the simple
  threshold sampling a Gaussian field \cite{kaiser} biasing is linear
  when fluctuations are small and Gaussian \cite{cdm_theo}. In
  addition only non-local biasing mechanisms, which at the moment have
  not been explored in the literature, could possibly produce large
  scale density fluctuations of the kind observed  in the
  galaxy distribution.

\section{Discussion}  

In summary, by applying the SL method to the 2dFGRS samples we detect
large density fluctuations of considerable spatial extension.  At
scales $r \ltapprox 40$ Mpc/h we find statistically stable power-law
correlations with fractal dimension $D=2.2 \pm 0.2$ in agreement with
previous determinations
\cite{slmp98,book,hogg,dr4_paper,2df_paper,paper1}.  For $r>40$ Mpc/h
we find that the galaxy distribution is strongly inhomogeneous and
fluctuations are large up to the samples sizes, in agreement with a
similar analysis of the SDSS data \cite{paper1}.  Persistent large
scale density fluctuations are compatible \cite{book} with fractal
power-law correlations extending to scales $r>$ 40 Mpc/h but
incompatible with homogeneity at $\lambda_0 \le 75$ Mpc/h. On the
other hand, standard models of galaxy formation, normalized to CMBR
anisotropies, predict $\lambda_0^m \approx 10$ Mpc/h
\cite{pee80,springel05}, i.e.  smaller than our lower limit $\lambda_0
> 75$ Mpc/h.  This prediction is in agreement with the results we
found in mock galaxy catalogs where we measured that, fluctuations are
more smoother than in the 2dFGRS samples, and their PDF rapidly
converges to a Gaussian function for $r>10$ Mpc/h.

Our results are in contrast with the standard determinations that the
characteristic length scale of galaxy distribution, marking the
transition to the regime of small fluctuations, is of the order of
$10$ Mpc/h \cite{zehavietal02,zehavietal04,norbergxi02}. This is
because this length scale is derived by measuring the amplitude of
two-point correlation function $\xi(r)$.  When considering this
quantity, which is normalized to the estimation of the sample density,
it is implicitly assumed that the distribution is homogeneous
(i.e. with small amplitude fluctuations) well inside the sample
volume, i.e. $\lambda_0 \ll V^{1/3}$ \cite{book}. When fluctuations
are large, as in the case of the 2dFGRS samples, this descriptor is
systematically biased by finite size effects
\cite{book,2df_paper,paper1} and so is the characteristic length scale
derived from its amplitude. On the other hand our results fairly agree
with studies of galaxy counts as a function of the apparent magnitude
$N(m)$, which indirectly probe radial distance fluctuations. These
show large fluctuations around the average behavior: particularly
$N(m)$ in the SGC are down by $30\%$ relative to the NGC counts
\cite{busswell03}.  These behaviors can be now directly related to
large scale galaxy structures and particularly to the fact that in the
NGC samples there are more structures, and thus an higher amplitude of
$\overline{N(r;R,\Delta r)}$ (Fig.\ref{fig_NR}), than in SGC
samples.

Finally it is worth noticing that our results agree with the
conclusion of \cite{einasto} who found that large scale structures
(e.g. super-clusters) are more frequent in observed samples than in
the simulations.

\acknowledgments We thank M. Joyce, A. Gabrielli, L.  Pietronero for
discussions and M. Blanton and M. Lopez-Correidoira for
comments. Y.V.B. thanks the partial financial support by
  Russian Federation grants ``Scientific school'' and ``Russian
  Education''. We acknowledge the use of the 2dFGRS data
  \cite{colless01} and of the Millennium simulations \cite{cronton06}.

\end{document}